\documentclass{article}
\usepackage{authblk}
\usepackage[top=2cm, bottom=2cm, left=2cm, right=2cm]{geometry}
\usepackage{fancyhdr}
\usepackage{setspace}
\usepackage{paralist}
\usepackage{hyperref}

\begin{document}

\title{A Canonical Password Strength Measure}
\author[1]{Eugene Panferov}
\date{} 

\maketitle

\begin{abstract}
\noindent We notice that the ``password security'' discourse is missing the most fundamental notion of the {\it password strength} -- it was never properly defined.
We propose a canonical definition of the {\it password strength}, based on the assessment of the efficiency of a set of possible guessing attack.
Unlike naive password strength assessments our metric takes into account the attacker's strategy, and we demonstrate the necessity of that feature.
This paper does NOT advise you to include ``at least three capital letters'', seven underscores, and a number thirteen in your password.
\end{abstract}
\vspace{2pc}

\section{Introduction}

We are constantly told to choose strong passwords -- but what is ``strong"?
A strong password is a password that is difficult to guess -- but what is ``guess" and what is ``difficult"? 
The present paper is trying to answer these two questions.

In order to define ``guess" we introduce a formal model of \emph{guessing attack}.
And then, in order to define ``difficult", we do the following steps: 
\begin{itemize}
\item we prove the theorem: \textbf{any two guessing attacks differ \underline{only by the order} in which they try candidate-passwords};
\item we demonstrate that ``password strength" (in any practical sense) is a function of an attack;
\item corollary the strength of a given password is the position of this password in the attacker's dictionary (because it is the only property this pair has);
\item the defender's strategy is represented by an approximation of the attacker's dictionary order;
\item easy to see that an approximate order is equivalent to a specific set of orders (i.e. different attacks);
\item thus, the defender's password strength is an expected value for the ``password strength" over the given set of attacks.
\end{itemize}

We offer to the reader a non-orthodox perspective: the password's strength is NOT a characteristic of a password, but a characteristic of an attack.
Mark Stockley (perhaps summarizing the most popular ideas on password security) writes \cite{stockley}:
{\it The next best option is to try to work out what characteristics passwords that are difficult to crack share}.
In our opinion this is exactly the pitfall that has trapped the majority of mainstream security experts.
Following this recommendation, whatever measure you create for (supposedly) the strength of the password,
it will be a measure of mimicry between passwords -- very likely irrelevant to the problem.  
The defender's strategy is not to mimic a guru's password, but to figure out a whole class of feasible attack strategies
and infer the properties they have in common, and then to create a password that is likely to defeat all these strategies.
The defender's strategy is to make reasonable attack strategies (which a reasonable attacker can not avoid) ineffective,
while keeping effective strategies unreasonable (which are very unlikely to be chosen by an attacker).

We focus on the one particular problem: making a password less susceptible for \emph{guessing}.
Thus, we assume that the attacker has no ``back door" access to the defender's hardware or software at all.
The attacker is trying to crack the password not the software nor hardware.
All he can use is a public (supposedly invulnerable) interface that expects a proper password to be input,
and outputs {\it success/failure} value immediately. The interface does not impose any limits on the amount of requests.
{\footnotesize This is equivalent to the ``offline guessing" attack \cite{schn}.}

Some important issues that are left out of the scope of the article: what if the attacker has a solid statistics
on the password usage virtually worldwide? what if the attacker is a service provider with his own massive password dictionary?
would it be a viable defensive move on the side of the defending service provider to publish his dictionary
or password usage statistics in order to devaluate the attacker's knowledge? --
The present paper provides a formal basis for answering these questions.

Please, note: by the word ``word" we do not refer to any natural language, but a finite sequence of symbols of a finite alphabet;
by the word ``dictionary" we do not refer to any natural language dictionary, but an ordered set of words.

\section{State Of The Art}

Despite the importance of the security issue in the ever growing world of electronic communications state of the art is as grim as medieval medicine.
Bruce Schneier has described the global conditions as:
``there's been a lot written on this topic over the years -- both serious and humorous -- 
but most of it seems to be based on anecdotal suggestions rather than actual analytic evidence'' \cite{schn}. 
The recent work \cite{zipf} clearly states the problem:
``Surprisingly, as far as we know, existing literature has not provided a satisfactory answer
to the above question of how to accurately measure the \emph{strength of a given password dataset}.''
And a popular website \cite{explained} illustrates these statements by giving a spectacularly hollow non-definition: ``Password strength is a measure of the effectiveness of a password in resisting guessing and brute-force attacks. In its usual form, it estimates how many trials an attacker who does not have direct access to the password would need, on average, to guess it correctly. The strength of a password is a function of length, complexity, and unpredictability.'' -- it obscures the matter it addresses in almost every word: ``effectiveness", ``resisting", ``complexity", ``unpredictability" -- what are they? 

The missing definition, however, does not stop the service providers from condemning ``weak'' passwords,
therefore, nearly all password creation policies are numerology, alchemy, and homeopathy such as the following examples:

\vspace{.5pc}

Google says \cite{google}:
{\it
Tips for creating a secure password:

\begin{compactitem}
\item{    Include similar looking substitutions, such as the number zero for the letter 'O'}
\item{    Create a unique acronym}
\item{    Include phonetic replacements}
\end{compactitem}
}

\vspace{.5pc}

Microsoft says \cite{ms}:
{\it
DO NOT USE:

\begin{compactitem}
\item{    Common letter-to-symbol conversions, such as changing ``o" to ``0"}
\item{    Abbreviations}
\item{    Common misspellings}
\end{compactitem}
}

\vspace{.5pc}

The same source:

{\it Give passwords the thought they deserve, and make them memorable. One way is to base them on the title of a favorite song or book, or a familiar slogan... }
-- {\footnotesize That's exactly what I would never do! That is a secure way to win the popularity contest for passwords
(according to \cite{schn2} 5th place belongs to ``blink182").
Once you refer to anything like a slogan you end up in the same bucket with over 9000 creative and unique other individuals watching the same commercials on TV. Besides that, \emph{The Titles} is a small dictionary, perhaps smaller than those ``cybercriminals' dictionaries'' referred to with awe and fear in the same paper. The link between the dictionary size and the password strength is the major topic of the next section.} 

\vspace{.5pc}
 
Another example is numerous online password-strength-meters (which could be recognized (by easy googlability) as representatives of the mainstream culture).
They do unanimously value the word ``P1ayer" higher than the ``letseatsomeofthoseprettygreenapplesdude".
Mark Stockley has made a good point on this topic \cite{stockley}.
Also \cite{zipf} emphasize the contradiction between ``strength'' values given by the variety of strength-meters to the same word.
Of course they contradict each other! What else could we expect in the absence of the very definition of the strength!

\vspace{.5pc}

In contrast to corporate and public guidelines and regulations, some papers seek to analyze password weaknesses empirically
\cite{zipf} \cite{weir} \cite{amico} \cite{schn2} \cite{wu} \cite{florencio}.
Bruce Schneier gave us a comprehensive overview of the popular dictionary attacks \cite{schn}. 

These papers unanimously agree that the widespread ``best practices'' have very little positive impact on security.
And we want to comment that this is not an unexpected result. There are total $|A|^n$ words of the length $n$ over the alphabet $A$.
Obviously, the effect of increasing the length $n$ dwarfs the effect of extending the alphabet $A$. 
In spite of that, all the ``best practices'' are strictly focused on the extending the alphabet (adding caps, numbers, etc). Even the U.S. Department of Homeland Security believes in the magical power of capital letters \cite{uscert}.
The following simple example reveals the futility of those ``best practices''.
Let's take a password vulnerable to a dictionary attack (i.e. a word of a natural language).
There are approx 23000 English 7-letters words commonly used by scrabble players (a very good mark for a dictionary attack).
If, according to the widespread recommendations, we require a capital letter the search space will be multiplied at very best by the factor 127.
It is next to nothing! And it is the most optimistic estimation.
In practice people will use just first capital or all capital in approx. 89\% \cite{weir}
which roughly means that the search space has increased merely by factor of 2.
In contrast to this, if we ask users to use two words instead of one, then in the same terms of the search space,
we get the factor 23000 -- without negative memorability impact, and without any practical clues for the attacker.
{\footnotesize The cardinality of the search space is not a password strength, but it is an optimistic estimate for one,
and a good measure for defender's strategy feasibility -- its low value helps us to sort out hopeless defender's strategies.} 
 
Also these papers present clear evidence (such as patterns in passwords) that the users are very good at mocking {\it proactive password checkers}.
Some of them \cite{wu} \cite{prost} \cite{amico} explicitly accent this point.
Indeed, the apparent absurdity of these checks alone (see above) is a strong motivation to mock them.

Besides that, a particularly valuable research \cite{weir} makes a good point about the false assertions upon which the ``best practices'' are based.
Matt Weir said about this work:
``Our findings were that the NIST model of password entropy does not match up with real world password usage or password cracking attacks. If that wasn't controversial enough, we then made the even more substantial claim that the current use of Shannon Entropy to model the security provided by human generated passwords at best provides no actionable information to the defender. At worse, it leads to a defender having an overly optimistic view of the security provided by their password creation policies while at the same time resulting in overly burdensome requirements for the end users.''

Still, Shannon's entropy is commonly used, and very popular web-services appear to be using it too.

Still, the very notion of strength remains undefined.

\vspace{1pc} 

Also, it is commonly asserted that password strength contradicts memorability.
No work is done to understand why is it so, if at all!
Every single paper on security asserts without a shed of evidence (quote from \cite{philosophical}):
``Human memory is limited and therefore users cannot remember secure passwords" -- what a leap of faith!
We tried to trace the origin of this fundamental assertion and failed.
This completely hollow claim is so widely accepted that no single paper attempted to however slightly substantiate the claim. 
On the same grounds we could assert that the mainstream security experts simply confuse strong passwords with ugly looking ones.

\section{The Problem}

We insist that there is no such objective quality of a password as {\it strength}.
What intuitively perceived as {\it password's strength} is a quality of a hypothetical attack against the password.
Indeed, if we choose a password ``Aj\^P\_(k13]*f9Ye'' it will be cracked in no time by any attacker 
who have a dictionary of candidate passwords consisting of the single word ``Aj\^P\_(k13]*f9Ye''.
Now, you are going to shout: ``Who on Earth in his right mind would ever do such a cracking attempt?!"
-- And it is exactly an argument from an attacker's strategy!
You can not define {\it password's strength} in isolation of a hypothetical attack,
and you do not agree to use just any attack strategy (as that one proposed above),
you want to match your password against a ``reasonable'' and ``probable'' and ``plausible'' strategy.
So, in order to estimate your password's strength, you have to make the theory of attacker's mind. 

In the following sections we will try to formalize the class of {\it guessing attacks},
and figure out what the attacker knows and what he can do in order to maximize his success.
Through the understanding of the attacker's strategies we can judge their efficiency against a particular password,
thus determining {\it password's strength}.

\vspace{1pc}

The {\it password's strength} is an estimate of the {\it attack's cost} from the defender's standpoint.

\subsection{The Model Of The Attack}

Please, note: by the word ``word" we do not refer to any natural language, but a finite sequence of symbols of a finite alphabet;
by the word ``dictionary" we do not refer to any natural language dictionary, but an ordered set of words.

Let's assume the attacker has a dictionary (essentially an ordered set) of all candidate passwords that is guaranteed to contain the unique desired password. 
We call for short all candidate passwords \emph{words} (elements of a dictionary), and the desired password \emph{the password}.
The attacker will check every word of the dictionary one by one for being the password until the check is successful.
The outcome of the check procedure is strictly binary (either {\it success} or {\it failure}), and immediate. 
We call this procedure a \emph{guessing attack}.

Easy to see that a guessing attack is finite and always successful.
Thus, we can characterize an attack by its length expressed in the number of iterations (number of words checked).
Shorter attacks are better for the attacker.
Apparently, the attack's length crucially depends on the ordering of the dictionary, so that the attacker's strategy includes the dictionary ordering.
Now we can show that the strategy consists of the dictionary ordering solely.

Immediately from the definition follows that the only information the attacker gathers during the attack
is described by a set of failed attempts: let's call it $D_{failed} \subset D$ where $D$ is the dictionary.
Let's assume we can reduce the dictionary according to the knowledge of $D_{failed}$, i.e. 
if $D_{failed}$ contains no password then $D_{improbable} \subset D$ contains no password.
On the other hand, $D$ contains ONLY ONE password, hence:
if $D_{failed}$ contains the password then $D_{improbable} \subset D$ contains no password.
Therefore \underline{unconditionally} $D_{improbable}$ contains no password --
we may simply remove $D_{improbable}$ from $D$ beforehand.
Thus, we return to the exact same starting point: A dictionary with only relation of linear order.
There are no other relations on the dictionary in any way connected to our problem.
we can decide about the order we will check and then remove elements, and we can not remove a single element unless we checked it.

Easy to see that a content difference between dictionaries is reducible to the reordering by the following trivial procedure: given two dictionaries $D_1$, $D_2$, append $D_1 \setminus D_2$ to $D_2$, and append $D_2 \setminus D_1$ to $D_1$; the attack cost is preserved because the dictionaries are guaranteed to contain the password.

\vspace{1pc}

\noindent {\sc The Golden Rule}:
The attacker's strategy is the way he orders his dictionary. There is nothing else he can do.
{\footnotesize It suggests us that delving into details of an attack, such as a word mangling method (which is a focus of many researchers) might be misleading. The most sophisticated attack differs from the stupidest attack by the dictionary order only.}

\vspace{1pc}

Now, as the attack's framework is known, the defender can theorize about the attacker's strategy.
Keeping in mind that the exact order of the attacker's dictionary $D$ can not be known to the defender,
we offer to ESTIMATE this order by a weak order $AS$, of the following form:

$$AS_1 \subset AS_2 \subset AS_3 \subset ... \subset AS_n,\ AS_n = D$$ 

Where the attack starts from $AS_1$ then proceeds to $AS_2$ etc.
It essentially means: we do not know in what order every subset $AS_i$ will be searched,
but we (to some extent) know that all elements of $AS_i$ must precede all elements of $AS_{i+1} \setminus AS_i$.
From the defender's perspective we refer to $AS$ as {\it attacker's strategy}.

Apparently the plausibility of {\it password's strength} derived from $AS$
entirely depends on how precise we estimate a real attacker's strategy.
In the section 3.3 we will discuss what we can do to this end.
Anyway, an attacker's strategy has now obtained a measurable form,
and we can define a measure that corresponds to the intuitive notion of {\it password's strength}.

\subsection{The Measure}

A {\it brute force attack} is a special case of the guessing attack where the dictionary is sorted randomly.
This seemingly absurd concept if applied to a subdictionary is very important to comprehend attacker's strategies,
and necessary to define {\it password's strength}. 

Given a guessing attack with a randomly sorted dictionary $D$
we can conceive a random variable {\it length of the attack} $L(D)$ which is the number of iterations before success.

Probabilities of success on each iteration $P_{i}$ where $i = 1,2,3...|D|$ are:
$$
P_1 = \frac{1}{|D|}, \quad
P_2 = \frac{|D|-1}{|D|} \times \frac{1}{|D|-1}, \quad
P_3 = \frac{|D|-1}{|D|} \times \frac{|D|-2}{|D|-1} \times \frac{1}{|D|-2}, \quad
...
$$
Easy to see that all these $P_i$ are equal to $|D|^{-1}$
therefore the Expected value
of the $L(D)$ for an unsorted dictionary D is:

$$
E(L(D)) = \sum_{i=1}^{|D|}{i \times P_i} = \frac{1+|D|}{2}
$$

This expected value estimates the cost of a brute force attack with the dictionary $D$
let's name it $BF(D)$. 
From the attacker's standpoint it corresponds to the situation when we know nothing about defender's password choosing habits
(besides the fact (or maybe assertion) that the password is in the dictionary)
so we resort to the {\it brute force}.
From the defender's standpoint it corresponds to the situation when we know (or reasonably estimate) the attackers dictionary,
but unable to figure out the order of the dictionary.

Using the definitions above we can finally introduce the {\it password strength}.
Given a password $pw$ and a strategy $AS$:
$$
AS:\ AS_1 \subset AS_2 \subset ... \subset AS_n 
$$
the {\it strength} of $pw$ against $AS$ is defined as:
$$
S(AS,pw) = |AS_{k-1}| + BF(AS_{k} \setminus AS_{k-1}), \quad \textrm{where} \quad pw \in AS_k \setminus AS_{k-1}.
$$

Note that in the case of brute force {\it password's strength} is literally equal to {\it attack's cost}.

\subsection{The Attacker's Strategy}

The defender's password choosing strategy is entirely based on the estimation of the attacker's strategy.
Therefore the attacker's strategy deserves a thorough investigation,
particularly interesting are ``moves" that an attacker can not avoid,
all those subdictionaries that ought to be checked.
The defender in turn should avoid picking elements from those dictionaries,
trying to avoid as many ``earlier" components of an attacker's strategy as possible
(NOTE: a dictionary (or a strategy component) is not necessarily defined by listing all its constituents).

Conversely, any comprehensive research on attacker's behavior improves our {\it password's strength} measure.
It allows the reader to reassess his password's strength by merely reading news instead of being cracked.

Although the goal of the present paper is far from giving you an unbeatable defender's strategy,
we want to discuss some ``easy prey" qualities of the attacker's strategies,
giving the reader some clues to a plausible attacker's strategy.

Remember, the attacker's strategy is an order on the universum of passwords.
Informally speaking, any attacker must arrange an attack so that more popular passwords
would be checked before less popular ones.

\vspace{1pc}

To this end we may consult statistical properties of the leaked password lists (such as \cite{} \cite{}).
It doesn't mean we are going to list all known passwords, we need to extract some 
distinctive properties, telling us: ``passwords with the property is more popular than passwords without it".

\vspace{1pc}

Another starting point for understanding attacker's strategies is the offline password crackers
(so called ``password recovery tools" -- enjoy the euphemism).

The same observation was done in \cite{weir}, here is the quote:
``In [the previous] paper, we designed a password cracking program that was trained on previously disclosed passwords.
Our current version of this cracking program learns information such as the frequency people use certain words,
case mangling, basic password structure, the probability of digits and special characters, etc.
and uses that information to construct a probabilistic context free grammar that models how people select passwords.
Our password cracker then proceeds to make guesses in probability order according to that grammar.
... We originally designed our cracker for law enforcement to help them deal with strong encryption,
but we quickly found out that it was also useful for the defender to give an estimate on how strong a password actually was,
or at least how different it was from the grammar that the password cracker was trained on.''
The paper they refer to is \cite{weir2}.

Many of those password crackers are described well enough to understand their strategies \cite{schn},
and the dictionaries for them are not top secret either (e.g. famous dic-0294).

\vspace{1pc}

Finally we can ask some questions about particular qualities of the strategies,
such as: shall we check the shorter words prior to the longer ones,
even though knowing that the users are advised to avoid short passwords?
Here is our half-answer to the last question.

Let there be a dictionary, from which the defender chooses a password, partitioned in two subsets:
$$
D = D_s \cup D_l,\ \textrm{where}\ D_s \cap D_l = \emptyset,\ |D_s| < |D_l|
$$
There are two mutually exclusive strategies on this dictionary:
$$
AS_s:\ D_s \subset D \quad\ \textrm{and}\ \quad AS_l:\ D_l \subset D
$$
Which one costs less?

Although we have limited the partition by only two subsets,
nevertheless it does not limit the scope of our deduction,
because the partition is very general, we can apply the following
deduction iteratively to as many different arbitrary partitions as we need.

Let's denote $P$ the probability that the defender's password $pw$ falls into $D_l$
and $1-P$ is the probability of the opposite outcome.

Let us calculate conditional {\it attack's cost} for each of the two 
possible defender's choices and each of two strategies separately:
$$
Cost(AS_s,pw\in D_s) = BF(D_s) $$ $$
Cost(AS_s,pw\in D_l) = |D_s| + BF(D_l) $$ $$
Cost(AS_l,pw\in D_s) = |D_l| + BF(D_s) $$ $$
Cost(AS_l,pw\in D_l) = BF(D_l) 
$$
Since all these are conditional expected values we summarize them according to the rules and get unconditional Costs of both strategies.
$$
Cost(AS_s) = \frac{1+|D_s|}{2} \times P + (|D_s| + \frac{1+|D_l|}{2}) \times (1 - P) $$ $$
Cost(AS_l) = (|D_l| + \frac{1+|D_s|}{2}) \times P + \frac{1+|D_l|}{2} \times (1 - P)
$$
Let's solve the inequality $Cost(AS_s) > Cost(AS_l)$ over $P$ 
$$
P < \frac{|D_s|}{|D|}
$$
Unsurprisingly the ``try longer first" strategy costs less
if and only if the defender chooses a password from the shorter dictionary with 
the probability lower than $\frac{|D_s|}{|D|}$ 
which is exactly the probability of the same outcome being randomly produced from $D$.
Because of that, if the defender however slightly tends to choose a password from $D_s$,
then the strategy ``shorter first" should be preferred.

This result is applicable to the situation where the defender is choosing a randomized password of arbitrary length.
If there is a choice whether to add yet another letter to the password or not,
then the threshold probability will be $\frac{1}{1+|A|}$ where $A$ is the alphabet. 

It is very unlikely the defender chooses randomly one element out of $D$,
more likely he chooses an arbitrary length first. 
In this case highly implausible that the shorter password probability is any less than $\frac{1}{1+|A|}$.

On the other hand, let us estimate the cost of a mistake, by comparing the worst case of $AS_s$
versus the best case of $AS_l$, i.e. when $P=0$
$$
Cost(AS_s) = |D_s| + Cost(AS_l)
$$
If $D_s$ and $D_l$ are sets of all words of the length $m$ and $m+1$ respectively then:
$$
\frac{Cost(AS_s)}{Cost(AS_l)} = 1 + \frac{1}{|A|} 
$$
The cost of a mistake is very low. Therefore the answer to the question is:
the shorter passwords are better to be checked before the longer ones (in the absence of other factors).

Surprisingly, this reasoning holds even if the defender chooses words from a natural language dictionary.
In this case the alphabet $A$, over which the defender composes his password, is the natural language dictionary itself.
And the length of the password is $1$.

Furthermore, we conjecture that the intuitive assertion that the attacker's alphabet is ``Latin'', ``alphanumeric'', or ``ASCII'',
or identical to the defender's alphabet is a primary source of panic in face of dictionary attacks (see section 2).
Same words could be produced from different alphabets.
Indeed, a password ``P1ayer'' is also a word over the alphabet $\{0,1\}$ of the length 48,
despite your mind uses another alphabet to produce it.
A dictionary attack is just an extreme example of a very big alphabet, that reduces words (borrowed from a natural language) to the length 1.
An obvious remedy follows immediately: make a password longer -- 2,3,4 symbols instead of just 1,
thus bloating the search space to the \underline{power} of 2,3,4.

\vspace{1pc}

We offer you a basic attacker's strategy as follows:
all combinations of 1 then 2 symbols, \\*
then all the most popular passwords (this raises the question about inclusion of a known black list in a particular attack) \\*
followed by all 4-digit integers \\* 
followed by dic-0294 ordered by popularity with every word followed by its mangled versions \\*
followed by the 2 words combinations from dic-0294 with 1 of the words for colour \\*
same with all adjectives \\*
and then all 2 words combinations ordered by popularity of individual words \\*
followed by all random strings of 3,4,5... symbols.

This is just an example, by no means we try to give you the final answer to the password question and the meaning of life and stuff.

\subsection{The Defender's Strategy}

This section may seem unrelated to the paper's topic, but it is important to understand that we could not give you the following advice before we defined the {\it password strength}, everything we are about to say below makes any sence at all because of the definition above.

\vspace{1pc}

There are three problems that could be solved in one shot.

(1) An inherent flaw of a password creation policy is that it conveys plenty of information about the passwords to the attacker.
Since the ``best practices'' have become widespread, at very least an attacker may profit slightly
by removing intact words from his dictionary. 
More careful examination may give him better insights about the password choosing strategies of the defenders.
Some researchers recommend to prohibit the most popular passwords \cite{zipf} -- this could easily become a huge passwords leaking source,
if the active passwords are used for sampling. This could be exploited by the attacker, therefore it would.

(2) Poor memorability of heavily mangled passwords, which is sadly a requirement nowadays.

(3) A human brain can not conceive a random anything. This feature is well understood by the attacker and exploited (refer to any paper in the reference list) .

\vspace{1pc}

\noindent {\sc The Solution:} Make your password longer. How long? As long as a 5 to 10 words sentence. \\*
Make it ``Zeus's 2nd daughter rode a pony through the jungle".

\vspace{1pc}

It immediately solves all aspects of the memorability problem.
Firstly, something meaningful is always easy to remember,
actually creating sentences is exactly the technique of memorizing words!
This is how they teach you to memorize your passwords -- why not to save you one step?  
Secondly, it allows you to accommodate those password mangling symbols (enforced by stupid policies) in a harmless manner,
that does not damage memorability.
At the same time numbers and punctuation used in a sentence does not satisfy the attacker's expectations of you using these symbols.

Lastly, in a sentence like that, the mangling space alone delivers you 6 ``bits of security": (1,2) do you capitalize names, sentences, all words? (3,4) do you omit spaces, punctuation? (5,6) do you have a bad habit of omitting possessive, or articles? (Not counting the mangling methods that damage memorability)

A dictionary attack?

\section{Conclusion}

The more narrow definition you give to the set of the good passwords the worse these passwords become.
In the very instant you tell the world THE best password it becomes one of the worst.
The set of good passwords is normally defined as negation of the known set of bad passwords, and it should remain this way.
There is no such thing as ``the best practice of password choosing'', there are bad practices, and bad choices, and the only thing we can do is to avoid them.
The proposed measure contributes to this cause, it tells us: ``how many candidate passwords could be seen worse than the given password''.

An important difference to the entropy-based measures is that the entropy is based on the \underline{assumption} of ``possible outcomes'' \cite{shannon}. This assumption can be very accurate in many cases such as measuring the computer memory size, but it has absolutely no substance in the context of password choosing. The proposed measure is based on the assumption of the attacker's strategy, which has a profound physical meaning in the context of our problem, and allows for a fruitful investigation.

\vspace{1pc}

How will the environment change when progressively more passwords begin to lose the popularity contest?


\begin{thebibliography}{29}

\bibitem{weir}  
  Matt Weir, Sudhir Aggarwal, Michael Collins, Henry Stern,
  \emph{Testing Metrics for Password Creation Policies by Attacking Large Sets of Revealed Passwords},
  ACM,
  2010,
  \\* {\footnotesize \url{http://goo.gl/wqcX}}

\bibitem{amico}
  Matteo Dell'Amico, Pietro Michiardi, Yves Roudier,
  \emph{Measuring Password Strength: An Empirical Analysis}
  arXiv:0907.3402 [cs.CR],  
  2009,
  \\* {\footnotesize \url{http://arxiv.org/abs/0907.3402}}

\bibitem{zipf}
  Ding Wang, Gaopeng Jian, Xinyi Huang, Ping Wang,
  \emph{Zipf's Law in Passwords},
  ACM,
  2015,
  \\* {\footnotesize \url{https://eprint.iacr.org/2014/631.pdf}}


\bibitem{google}
  \emph{Google Accounts Help Center},
  Google,
  2015,
  \\* {\footnotesize \url{https://accounts.google.com/PasswordHelp}}

\bibitem{ms}
  \emph{Microsoft Safety And Security Center},
  Microsoft,
  2014,
  \\* {\footnotesize \url{https://www.microsoft.com/security/pc-security/password-checker.aspx}}

\bibitem{schn}
  Bruce Schneier,
  \emph{Choosing Secure Password},
  2007,
  \\* {\footnotesize \url{https://www.schneier.com/blog/archives/2007/01/choosing_secure.html}}

\bibitem{schn2}
  Bruce Schneier,
  \emph{MySpace Passwords Aren't So Dumb}
  Wired.com,    
  2006,
  \\* {\footnotesize \url{http://archive.wired.com/politics/security/commentary/securitymatters/2006/12/72300}}

\bibitem{florencio}
  Dinei Florencio, Cormac Herley,
  \emph{A Large-Scale Study of Web Password Habits},
  Microsoft Research,
  2007,
  \\* {\footnotesize \url{http://research.microsoft.com/pubs/74164/www2007.pdf}}

\bibitem{weir2}
  M. Weir, Sudhir Aggarwal, Breno de Medeiros, Bill Glodek,
  \emph{Password Cracking Using Probabilistic Context Free Grammars},
  Proceedings of the 30th IEEE Symposium on Security and Privacy,
  May 2009.

\bibitem{prost}
  S. Prost,
  \emph{A brief analysis of 40,000 leaked MySpace passwords}
  2007,
  \\* {\footnotesize \url{http://www.the-interweb.com/serendipity/index.php?}}
  \\* {\footnotesize \url{/archives/94-A-brief-analysis-of-40,000-leaked-MySpace-passwords.html}}

\bibitem{wu}
  Thomas Wu,
  \emph{A real-world analysis of Kerberos password security},
  Proceedings of 1999 Network and Distributed System Security Symposium,
  1999,
  \\* {\footnotesize \url{https://www.gnu.org/software/shishi/wu99realworld.pdf}}

\bibitem{stockley}
  Mark Stockley,
  \emph{Why you can't trust password strength meters},
  Naked Security,
  2015, 
  \\* {\footnotesize \url{https://nakedsecurity.sophos.com/2015/03/02/why-you-cant-trust-password-strength-meters/}}

\bibitem{philosophical}
  M Atif Qureshi, Arjumand Younus, Arslan Ahmed Khan,
  \emph{Philosophical Survey of Passwords},
  IJCSI International Journal of Computer Science Issues, Vol. 2,
  2009

\bibitem{shannon}
  Shannon, Claude E.,
	\emph{A Mathematical Theory of Communication},
	1948
 
\bibitem{uscert}
  U.S. Department of Homeland Security,
	\emph{Choosing And Protecting Your Password},
	2013,
	\\* {\footnotesize \url{https://www.us-cert.gov/ncas/tips/ST04-002}}

\bibitem{explained}
  everything.explained.today,
	\emph{Password Strength Explained},
	\\* {\footnotesize \url{http://everything.explained.today/Password_strength/#Ref-1}}

\end{thebibliography}
\end{document}